\documentclass[sigconf, pbalance=true]{acmart}

\begin{document}


\setcopyright{none}
\settopmatter{printacmref=false, printccs=false, printfolios=true}
\renewcommand\footnotetextcopyrightpermission[1]{}
\addtolength{\textheight}{9pt}

\title{A Periodic Space of Distributed Computing: Vision \& Framework}

\author{Mohsen Amini Salehi}
\authornote{These authors contributed equally to this work.}
\affiliation{%
  \institution{University of North Texas}
  \city{Denton}
  \state{TX}
  \country{USA}}
\email{mohsen.aminisalehi@unt.edu}

\author{Adel N. Toosi}
\authornotemark[1]
\affiliation{%
  \institution{The University of Melbourne}
  \city{Melbourne}
  \country{Australia}}
\email{adel.toosi@unimelb.edu.au}

\author{Hai Duc Nguyen}
\authornotemark[1]
\affiliation{%
  \institution{Argonne National Laboratory}
  \city{Lemont}
  \state{IL}
  \country{USA}}
\email{hai.nguyen@anl.gov}

\author{Murtaza Rangwala}
\authornotemark[1]
\affiliation{%
  \institution{The University of Melbourne}
  \city{Melbourne}
  \country{Australia}}
\email{mrangwala@student.unimelb.edu.au}

\author{Omer Rana}
\affiliation{%
  \institution{Cardiff University}
  \city{Cardiff}
  \country{UK}}
\email{RanaOF@cardiff.ac.uk}

\author{Tevfik Kosar}
\affiliation{%
  \institution{University at Buffalo}
  \city{Buffalo}
  \state{NY}
  \country{USA}}
\email{tkosar@buffalo.edu}

\author{Valeria Cardellini}
\affiliation{%
  \institution{Tor Vergata University of Rome}
  \city{Rome}
  \country{Italy}}
\email{cardellini@ing.uniroma2.it}



\author{Rajkumar Buyya}
\affiliation{%
  \institution{The University of Melbourne}
  \city{Melbourne}
  \country{Australia}}
\email{rbuyya@unimelb.edu.au}

\begin{abstract}
Advances in networking and computing technologies throughout the early decades of the 21st century have transformed long-standing dreams of pervasive communication and computation into reality. These technologies now form a rapidly evolving and increasingly complex global infrastructure that will underpin the next aspiration of computing: supporting intelligent systems with human-level or even superhuman capabilities.
We examine how today’s distributed computing landscape can evolve to meet the demands of future users, intelligent systems, and emerging application domains. We propose a ``periodic framework'' for characterizing the distributed computing landscape, inspired by the systematic structure and explanatory power of the ``periodic table'' in chemistry.  This framework provides a structured way to describe, compare, and reason about the behaviors and design choices of different distributed computing solutions. Using this framework, we can identify patterns in key system properties, such as responsiveness and availability, across the distributed computing landscape. We also explain how the framework can help in predicting future trajectories in the field.
Lastly, we synthesize insights from leading researchers worldwide regarding the desired properties, design principles, and implications of the emerging areas in the forthcoming distributed computing landscape and in relation to the periodic framework. Together, these perspectives shed light on the considerations that will shape the distributed computing landscape underpinning future intelligent systems.

\end{abstract}


\maketitle
\fancyhead[RO,LE,RE,LO]{}

\section{Introduction}
\label{sec:intro}

The first decades of the 21st century witnessed rapid progress toward realizing two long-standing aspirations of computing: pervasive communication and computation. Successive generations of networking technologies, the widespread adoption of Internet-connected devices, and more recently low-Earth-orbit satellite constellations \cite{lagunas2024leo,ma2023network} have transformed connectivity from a regional capability into a truly global one. In parallel, hyperscale cloud platforms \cite{Armbrust2010Cloud} and the proliferation of IoT systems \cite{Atzori2010IoT} have reshaped computing into a continuum of interconnected and heterogeneous tiers spanning devices, edge and fog layers, and large-scale cloud infrastructures, each characterized by distinct constraints and service abstractions \cite{hotmobile19}. This evolution has enabled computation to migrate fluidly across the continuum to satisfy application requirements for latency, privacy, and quality of service \cite{globecome23,jewcloud24,Akbari2025}. As a result, distributed computing systems are transitioning from an active research domain into critical global infrastructure \cite{WhiteHouse2020CET,EC2020Digital}, forming the foundation for distributed intelligence \cite{Duan2023DAI}, where AI systems operate collectively across large-scale environments. Understanding the dynamics of this emerging landscape is therefore essential, as it will underpin future intelligent systems and may ultimately shape the technological trajectory of the 21st century.

As computing environments become increasingly pervasive and interconnected, distributed systems are approaching a fundamental \textit{paradigm shift} in how they are designed, governed, and consumed. In this emerging landscape, computing resources are becoming a strategic asset, often compared to the ``oil'' of the 21st century. The rapid rise of infrastructure-intensive technologies such as generative AI and large language models (LLMs) \cite{Bommasani2021Foundation}, combined with the slowing of Moore’s Law as semiconductor scaling approaches physical limits \cite{Theis2017MooresLaw}, suggests that the coming decades will be defined by a profound bottleneck in computational capacity \cite{Thompson2021Decline}. This shift introduces both opportunities and challenges. Limited access to advanced computing infrastructure, particularly in developing regions \cite{UNCTAD2019Digital}, risks widening the digital divide and constraining productivity, innovation, and global competitiveness \cite{WorldBank2016WDR}. At the same time, the rapid expansion of distributed systems and hyperscale data centers brings significant societal and environmental costs, including rising energy and water consumption, noise pollution, and an enlarged attack surface for cyber threats \cite{bashir25}. Together, these trends highlight the growing importance of rethinking how future distributed computing infrastructures are structured and managed.

Understanding how distributed computing infrastructures should evolve in response to these pressures requires a careful examination of its \emph{characteristics}, \emph{evolution}, and \emph{ramifications}. Yet this task is increasingly difficult as the distributed computing continuum continues to expand in both scale and complexity, with new frameworks, architectures, and abstractions emerging at a rapid pace, each introducing distinct assumptions and design trade-offs. This rapid evolution makes it challenging to form a unified view of the ecosystem or to reason systematically about its long-term trajectory.
To address this challenge, we argue for a principled approach to organizing the computing landscape into a self-explanatory and predictive structure---one capable of revealing relationships, constraints, and system behaviors without requiring detailed knowledge of individual solutions. A similar intellectual breakthrough occurred with the development of the Chemical Periodic Table \cite{black2010review}, which organized elements in a way that made their properties and relationships interpretable and even enabled predictions of undiscovered elements. Inspired by this idea, we seek an analogous organizing principle: a conceptual \textbf{periodic space} that provides an intuitive and systematic framework for characterizing \textit{existing} and \textit{future} solutions across the distributed computing landscape.

Building on this principle, we propose a framework that captures the fundamental design constraints and system properties of distributed computing across the \textit{tiers continuum} and \textit{levels of abstraction}. We examine how this framework can describe recurring patterns and help reason about the behavior of existing and emerging solutions across the distributed computing landscape. To ground this perspective, we also incorporate insights from leading researchers working at the boundaries of the proposed periodic space. By providing a structured view of the evolving ecosystem, this work aims to inform researchers, technology leaders, and policymakers in shaping the next generation of distributed systems.

\section{Periodic Framework of Distributed Systems}
\label{sec:periodic}

\subsection{Periodic Space}

\begin{figure}[t]
    \centering
    \includegraphics[width=\linewidth, trim={0 0 0 0},clip]{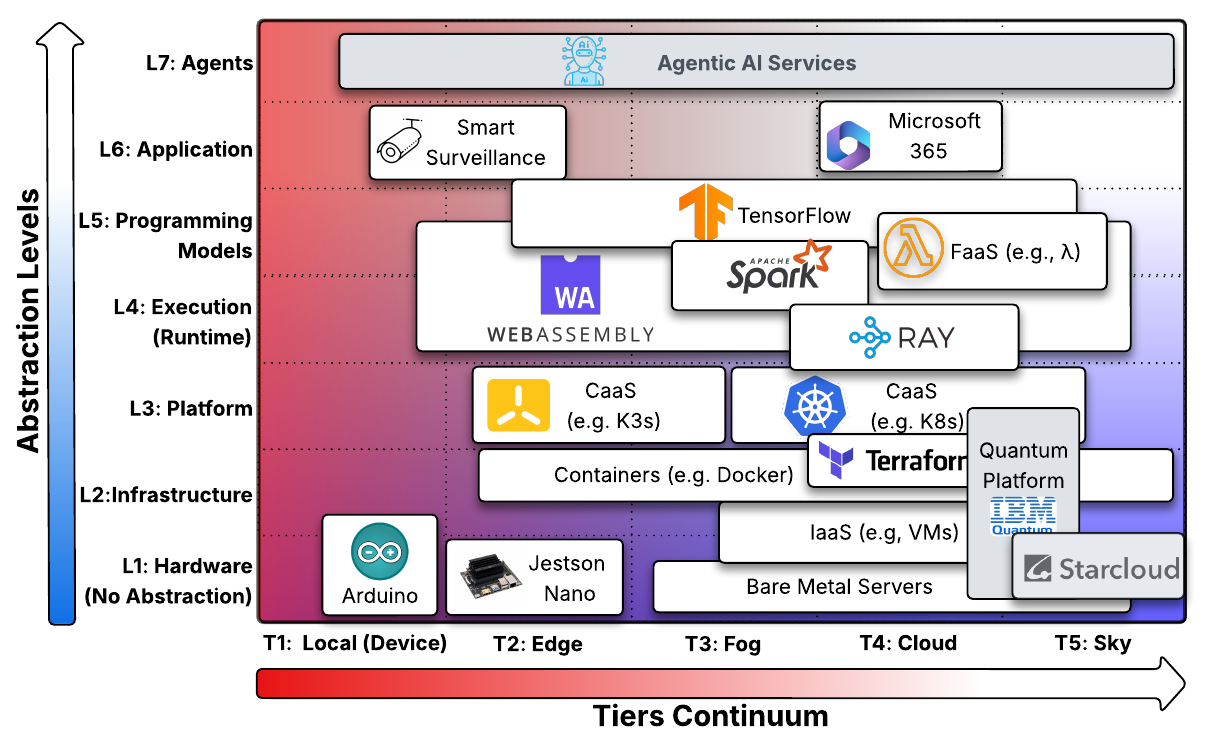}
    \caption{A ``Periodic Space'' of distributed computing landscape, organized by \textbf{Tiers} (horizontal) and  \textbf{Abstractions} (vertical). Representative solutions are positioned within the space. \textbf{Gray boxes} indicate futuristic or emerging solutions. 
    }
    \label{fig:periodic}
\end{figure}


We propose the concept of a \textbf{Periodic Space} for distributed computing, illustrated in Figure~\ref{fig:periodic}. The central idea is to provide a conceptual framework that organizes the distributed computing landscape into a continuous and extensible space, enabling solutions to be characterized by their underlying constraints, capabilities, and levels of abstraction rather than by isolated taxonomies or niche properties. In doing so, the periodic space offers a systematic way to describe relationships among existing solutions while remaining flexible enough to accommodate emerging solutions as the distributed computing ecosystem continues to evolve.

This framework arranges computing solutions along two orthogonal dimensions: \textbf{tiers continuum} on the horizontal axis, spanning \textit{computing infrastructure} from the extreme local devices and edge, to centralized cloud and beyond; and the \textbf{abstraction level} on the vertical axis, capturing the progression from low-level hardware to high-level Agents AI services. By positioning existing solutions within this space, the framework provides a unified lens for explaining relationships among solutions, highlighting design trade-offs, and identifying unexplored or emerging regions that point toward future computing models. We note that the chemical periodic table is discrete, i.e., each element has a single position in the table that reflects its fundamental properties and relationships (e.g., atomic structure and chemical behavior). However, our \emph{Periodic Space} represents a continuous space where a given solution may span over more than one tier and abstraction level.


\subsubsection{Tiers Continuum}
\label{sec:periodic-resource-coverage}

To support applications with diverse requirements in performance, reliability, and data locality, computing infrastructure has naturally evolved across multiple physical layers. At one end of the spectrum are resource-constrained devices (e.g., IoT sensors and cameras) where computation occurs close to data sources. At the other end are hyperscale cloud data centers designed for large-scale analytics and elastic resource provisioning. These layers reflect fundamental trade-offs among proximity to data, resource capacity, and operational scope.
Positioning distributed computing solutions along this dimension, based on \textit{where} their underlying resources reside and the \textit{scale} at which they operate, provides a natural way to reason about, compare, and understand their relationship with the underlying infrastructure. In practice, there are no strict boundaries between computation tiers, so we make the spectrum infinite and continuous to allow future expansion. However, for clearer presentation and understanding, we use commonly recognized tiers as reference anchors across this continuum.

 \noindent \textbf{Local Device}---Computation occurs at the same location where data is generated or user requests originate (e.g., on-device video analytics \cite{huynh2017deepmon} and sensor data processing on wearables \cite{yeon2025watchhar}).
    
\noindent\textbf{Edge}---Computation is placed within the same local network or physical vicinity as the data source, such as servers deployed in factories \cite{nain2022towards} or mobile and energy-constrained systems \cite{satya25, mokhtari2024heet}.

\noindent\textbf{Fog} (aka Cloudlet)---resources are provided by regional data centers with no energy constraint, some level of elasticity, and slightly more latency in compare to the edge tier \cite{charyyev2020latency}. Examples are campus clusters and ISP-operated mini–data centers that offer domain-specific services~\cite{felare23} . 
    
\noindent\textbf{Cloud}---resources come from large, widely distributed (public or private) datacenters and can be provisioned globally. Clouds are characterized by their elasticity, and large-scale computation for use cases such as big data analytics, and machine learning training. Major global public cloud providers are AWS, Azure, and GCP.

\noindent\textbf{Sky}---resources extend beyond a single cloud to encompass cross-cloud deployments~\cite{sky21}. More recently, this concept has been extended to computing infrastructures spanning terrestrial, aerial, and space-based platforms~\cite{liu2018space}, such as LEO satellites and high-altitude platforms, and even extra-planetary computing infrastructures, such as planetary-scale and off-world data centers \cite{denby2020orbital}.

\subsubsection{Abstraction Levels}
\label{sec:periodic-resource-abstraction}
As computing resources expand across the tiers continuum, developing, deploying, and maintaining distributed applications has become increasingly complex. To manage this complexity, software layers introduce abstractions that hide infrastructure details such as hardware heterogeneity, resource placement, communication, and failure handling, allowing developers to focus on application logic rather than the mechanics of distribution. Higher levels of abstraction simplify development and deployment but reduce visibility into, and control over, the underlying resources. Similar to the tiers dimension, we model abstraction as an infinite and continuous spectrum, since infrastructure can be abstracted in many ways across computation, communication, and data management without clear boundaries between levels. Because abstractions emerge from developer needs, distributed solutions can be positioned along this spectrum according to which parts of their software stack are abstracted away:

\noindent\textbf{Bare Hardware (No Abstraction)}---solutions expose raw hardware with no abstraction, such as bare metal servers in data centers, or naked single board computers at the edge (e.g., Raspberry Pi \cite{raspberrypi} or an Nvidia Jetson Nano \cite{nvidia_jetson_nano}).

\noindent\textbf{Infrastructure}---hardware complexity is abstracted through virtualization technologies such as virtual machines, containers, or emerging WebAssembly-based runtimes \cite{zhang2025research, menetrey2022wasm}. 

\noindent\textbf{Platform}---environment dependencies are abstracted away, including operating systems, libraries, and runtime configuration. By shielding developers from low-level infrastructure management, the platform layer enables consistent deployment and execution across heterogeneous resources. Cloud middleware platforms (e.g., Kubernetes \cite{jonas2019cloud}) or open-source Infrastructure as Code (e.g., Terraform \cite{moris2021infrastructure}) are categorized in this level.

\noindent\textbf{Execution (Runtime)}---provides a managed execution environment for running applications via abstracting operational tasks such as resource management, elasticity, and fault-tolerance. Examples include serverless execution environments (e.g., KNative \cite{knative}) that execute code in response to events.

\noindent\textbf{Programming Models} --- abstract away complexities at application scale (i.e., cross-solution integration), including workflow orchestration (AWS Step Functions \cite{aws_step_functions}), Quality of Service (QoS) guarantees (RBAM \cite{nguyen2025efficient, nguyen2019real}), end-to-end development and deployment (e.g., OaaS \cite{wissocc24,wistc25,wisipdps26}).

\noindent\textbf{Application} aka software as a service (SaaS)---in this level, a service provides the entire ready-to-use software application with minimal configuration (e.g., Microsoft 365 \cite{microsoft_office}).

\noindent\textbf{Agents}---abstract away human effort on developing, deploying, and managing solutions, currently through autonomous or semi-autonomous agentic services \cite{park2023generative} that encapsulate application logic, decision-making, and action execution. Examples include autonomous data analysis agents \cite{fu2025autonomous}, task-oriented copilots \cite{bird2022taking}, and multi-agent systems \cite{li2024survey}.


\subsection{Periodic Space for System Properties}


Just as the periodic table of chemical elements conveys fundamental properties based on the position of each element, our proposed periodic space can serve as a lens for understanding and characterizing solutions and systems within the space with respect to a range of secondary dimensions, which we collectively refer to as ``\textbf{system properties}''. By tracing trends of these properties across the periodic-space framework, we enable structured comparison and obtain deeper insight into how placement across tiers and abstraction levels influences overall system behavior. In particular, this structure allows us to reason how a given solution is likely to behave as it moves along the periodic space.

\begin{figure}[h]
    \centering
\includegraphics[width=0.99\linewidth]{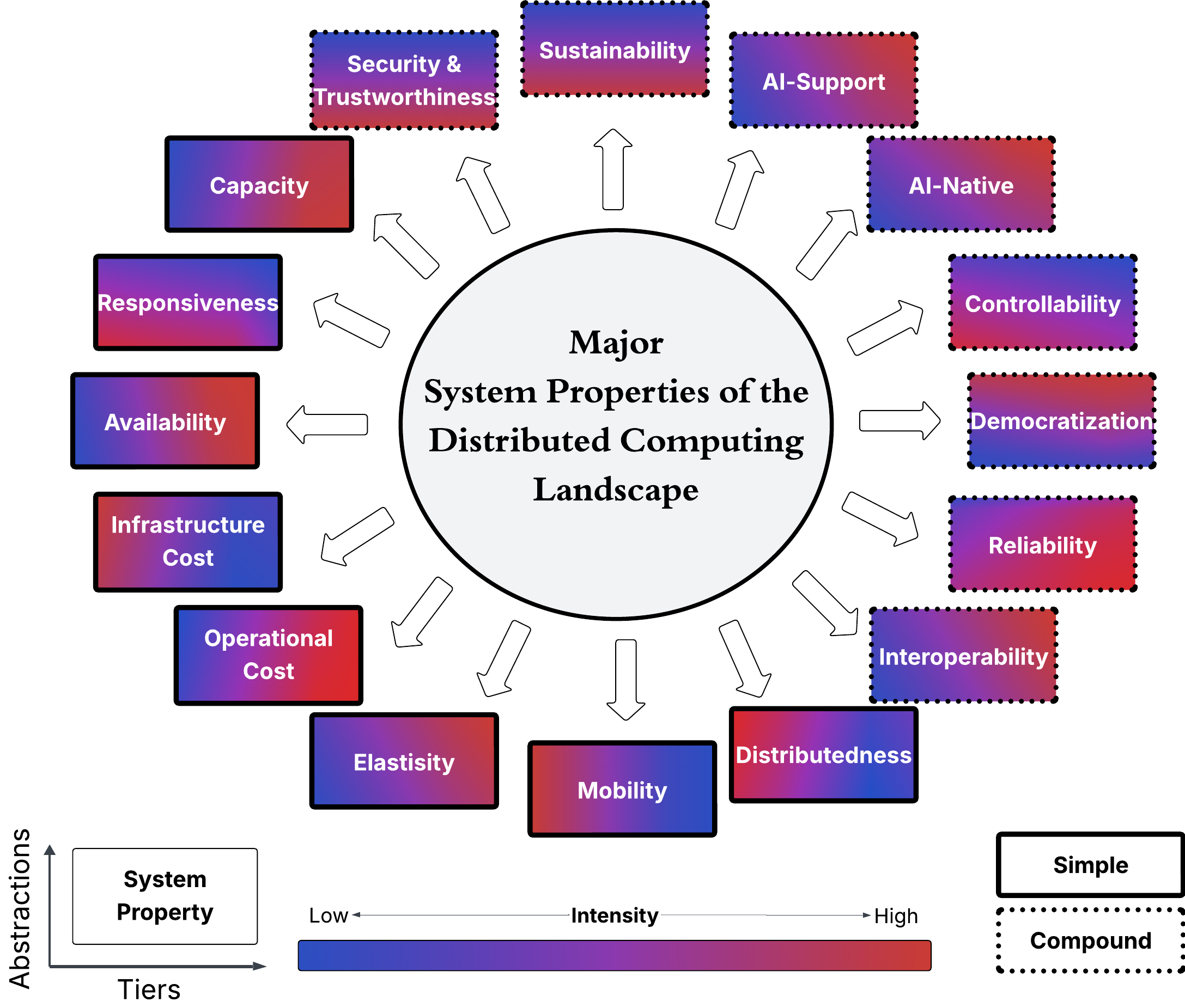}
    \caption{
    System properties across computing tiers (left to right: local → sky) and abstraction levels (bottom to top: hardware → agent), categorized as \textit{simple} and \textit{compound}; color indicates intensity relative to the periodic space.
    }
    \label{fig:syspro}
\end{figure}

We identify a set of key system properties, summarized in Figure~\ref{fig:syspro}, with their definitions provided in Table~\ref{tab:system-properties}. These properties often exhibit distinct, either uniform or non-uniform trends across the periodic space. We categorize the properties into two groups: \textit{Simple} and \textit{Compound}. Simple properties, e.g., responsiveness and capacity, typically exhibit more straightforward trends across the periodic space and are easier to intuitively represent and reason about as the space evolves. In contrast, Compound properties emerge from the interaction of multiple simple properties, hence, require multi-metric evaluation. For example, reliability depends on latency, failure rates, and the degree of distribution, and governance depends on access control and observability.

\begin{table*}[h]
\caption{Key system properties of solutions in distributed systems; along with their definition and measurement (evaluation) metric. The last two columns show how each property is impacted by changes across the periodic space: $\uparrow$/$\downarrow$ = increase/decrease as abstraction/tier increase, $-$ = no impact, and $?$ = unknown, unclear or non-linear pattern$^{\star}$.
}
\label{tab:system-properties}
\begin{tabular}{p{0.15\linewidth} p{0.45\linewidth} p{0.15\linewidth} c c}
\hline
\textbf{Property} & \textbf{Definition} & \textbf{Common Measurement} & \textbf{Abstraction} & \textbf{Tiers}\\
\hline

\textbf{Capacity} &
Computational power a solution provides in terms of processing, memory, storage, and network.
& Throughput (requests/sec) & $-$ & $\uparrow$ \\

\textbf{Responsiveness} & Speed with which a solution processes and returns responses to user/client requests. & Latency (sec) & $\downarrow$ & $\downarrow$ \\

\textbf{Capital Cost}$^{*}$ & The total software \& hardware expenses to  develop and establish a solution. & Money (\$) & $\uparrow$ & $\downarrow$\\

\textbf{Operational Cost}$^{*}$ & The total economic expense to operate and run a solution. & Money (\$) & $\downarrow$ & $\uparrow$\\

\textbf{Elasticity} & The ability of a solution to adapt its resource allocation to load changes in order to maintain performance objectives. & Load-Metric Curve or Elasticity ratio& $\uparrow$ & $\uparrow$\\

\textbf{Reliability} & The ability of a solution to provide correct and dependable services over time.
& Availability (\%) & $?$ & $\uparrow$ \\

\textbf{Mobility} & The ability of a solution to work on mobile compute resources (e.g., smartphones or vehicular nodes) while maintaining uninterrupted services. & N/A & $-$ & $\downarrow$ \\

\textbf{Distributedness} & The extent to which computation, data, and control are spread across multiple geographically and logically distinct nodes.
& N/A & $\uparrow$ & $\downarrow^\dagger$\\ 

\textbf{Interoperability} & The ability of a solution to operate across platforms, technologies, and administrative domains. & N/A & $\uparrow$ & $?$\\

\textbf{Democratization} & The degree to which a solution lowers barriers to access, development, and deployment through ease of use, programmability, and accessibility. & Learning Hours or Time to Delivery & $\uparrow$ & $?$ \\

\textbf{Controllability} & The degree to which developers or users can directly configure, manage, and influence infrastructure behavior and execution decisions.
& N/A & $\downarrow$ & $\downarrow$\\

\textbf{AI-Native} & AI-optimized solutions where AI is fundamentally integrated into the system’s design and operation. 
& N/A & $?$ & $\uparrow$ \\

\textbf{AI-Support} & Solution that are built to efficiently support AI workloads. 
& N/A & $?$ & $\uparrow$ \\

\textbf{Sustainability$^{\ddagger}$} & The ability of a solution to minimize carbon emission while remaining effective over time. & Carbon emission ($CO_{2}e$) & $\downarrow$ & $\uparrow$  \\

\textbf{Security \& Trustworthiness} & The ability of a solution to protect data, operations, and users against threats while ensuring integrity, confidentiality, and trustworthy behavior. & N/A & $?$ & $?$ \\
\hline
\end{tabular}
\par\vspace{2mm}
\noindent\small $^{\star}$For more insights, please refer to the website we developed for this purpose: \url{https://hpcclab.github.io/periodic-table}. 
\noindent\small $^{*}$These metrics are to be interpreted from the service provider's point of view. 
\noindent\small $^{\dagger}$This trend will show a slight increase as it moves from the Cloud tier to the Sky Tier. 
\noindent\small $^{\ddagger}$This metric is strictly considered from the lens of energy usage. 
\end{table*}


As illustrated in Figure~\ref{fig:syspro}, we characterize each system property by highlighting its trends and variations across the periodic space. These trends and variations are visually encoded using intensity-based background coloring for each property. Through several examples below, we demonstrate how the periodic space can be used to characterize and reason about system properties across the distributed computing landscape.

\noindent Example 1: \textbf{Responsiveness.} As a solution moves from lower, locality-oriented tiers (e.g., a container at the edge) to higher tiers (the same container deployed in the cloud), it typically experiences increased latency and reduced responsiveness due to greater physical and network distance from end users and data sources \cite{satyanarayanan2017emergence}. 

\noindent Example 2: \textbf{Democratization} and \textbf{Ease-of-Use.} When a solution moves up across abstraction levels, it becomes easier to develop, deploy, and maintain--reflecting increased \textit{democratization} and \textit{ease-of-use}. However, this often comes at the cost of lower \textit{governance} and \textit{trustworthiness}, as control over underlying system behavior is increasingly delegated to the platform \cite{younis2024comprehensive}.

\noindent Example 3: \textbf{Elasticity.} This property exhibits clear trends across both tier and abstraction dimensions. Higher-level tiers, such as the cloud, naturally provide more elastic behavior than local or edge tiers due to their access to larger resource pools. In addition, introducing higher levels of software abstraction in the stack enables greater flexibility in resource management--allowing more elastic solutions in compare to deployments tightly coupled to non-abstracted hardware.

\subsection{Periodic Space to Anticipate Future Trends}

Beyond characterization, the periodic space also has an \textit{anticipative} role: by observing how system properties evolve along its dimensions, we can reason about likely future trajectories of the field. For example, as Moore’s law continues to slow down and hardware specialization accelerates \cite{shalf2020}, emerging computing substrates such as quantum processors and domain-specific accelerators are unlikely to appear uniformly across the periodic space. Instead, consistent with the trends observed in existing system properties, their high cost, operational complexity, and reliability requirements suggest that they will initially be positioned in higher-level computing tiers, most notably in cloud environments \cite{rallis2025interfacing, kumari2025quantum}. Moreover, to preserve desirable system properties such as democratization and ease-of-use while integrating these specialized resources, new abstraction layers will be required. These abstractions will shift solutions upward in the periodic space, enabling developers to exploit novel hardware capabilities without deep hardware-specific expertise. 

As we can see, the periodic space not only describes current solutions, but also can provide a structured lens to predict how future technologies will reshape system behavior across tiers and abstraction levels. 
To support deeper exploration of these trends for different solutions, we have prepared an interactive webpage\footnotemark[\value{footnote}]\footnotetext{Interactive periodic space: \url{https://hpcclab.github.io/periodic-table}}
 that allows users to select a property and visualize its behavior across the periodic space for a given solution.

\section{Emerging Distributed Systems through the Lens of Periodic Space}


\subsection{Overview: Agent Abstraction and Sky Tier} 
As both infrastructure and applications continue to grow, the periodic space will be expanded in both  abstraction and tier dimensions.

\textbf{Abstractions.} As resource capacity is continually increasing across all tiers, solutions that were traditionally confined to upper level tiers (e.g., cloud-based FaaS \cite{kounev2023serverless}) will be applicable at lower tiers (e.g., edge). However, increasing capability at the lower tiers also brings greater hardware heterogeneity and complexity, demanding for stronger abstractions in those tiers. For instance, Container as a Service (CaaS) \cite{burns2016borg} and WebAssembly (WASM) \cite{haas2017wasm} abstractions will be used at the ``device'' tier to democratize complex deployments \cite{kakati2024wasm}.

To manage the increasing complexity, we envision that the abstraction spectrum will also continue to grow towards: (a) new programming-model abstractions that unify access to diverse specialized hardware across multiple tiers \cite{wiscloud23,wistc25}; and (b) higher-level \textit{agent} abstractions, where developers and end-users can rapidly construct personalized applications on demand across the continuum using LLM-driven agentic systems \cite{esashi2026action}, instead of relying solely on fixed, off-the-shelf SaaS offerings. Agents operate above applications by encapsulating intent, goals, and adaptive behavior \cite{park2023generative}. They abstract away both application-specific implementation details \cite{CANLMAki26} and underlying system complexities. 

\textbf{Sky Tier.} On the tiers dimension, escalating computational demand, particularly from AI, will push the tiers upward---demanding for more interoperability across clouds and hyperscale supercomputing centers towards a ``Sky tier'' that can collectively meet unprecedented workload volumes. 
This tier represents a highly federated, globally distributed, and predominantly autonomous computing tier composed of multi-cloud, cross-domain, emerging non-terrestrial (orbital, lunar, and deep-space systems), and quantum computing resources \cite{sky24} 
to enable artificial superintelligence \cite{grabowska2024quantum} and other emerging workloads.


\subsection{Trends Towards the Sky Tier}

\subsubsection{Reliability} 
As distributed systems evolve toward the Sky tier, their scale approaches planetary levels, where infrastructure spans heterogeneous hardware, multiple administrative domains, and diverse physical environments. Network disruptions, hardware faults, software anomalies, and resource variability become inevitable at this scale, making failures no longer rare or exceptional events but continuous operating conditions \cite{nicolae2024diaspora,nguyen2025resilient}. Consequently, Sky-tier systems must assume failures as part of normal operation, and resilience mechanisms that enable systems to continue functioning seamlessly despite ongoing disruptions become an essential property. However, achieving such resilience becomes increasingly challenging as systems scale across tiers, where complexity grows along multiple dimensions, including heterogeneity, connectivity patterns, and software dependencies. This makes reliability difficult to maintain at low abstraction levels, such as individual devices or infrastructure components. Instead, reliability efforts must shift toward higher abstraction layers, where complexity is hidden behind high-level service guarantees and simplified operational models.


\subsubsection{Fluidity}
The future Sky tier will demand abstractions to enable truly seamless interoperability (aka fluidity), such that computation can migrate as seamlessly as data packets do today \cite{jewcloud24}. Such fluidity should support both vertical migration, across edge-cloud tiers, or horizontal, across multi-cloud. As an example of vertical fluidity consider a group of youth playing game on a coffee shop edge server. Upon arrival of a disabled customer, the game process should seamlessly migrate to cloud to free resources for the disabled user; migration in the opposite direction can happen once the user leaves the place \cite{globecome23}. Horizontal fluidity, however, is getting popular across space datacenters, such as those in low-Earth orbit to maintain services deployed in a certain geographic region~\cite{kometsocc24,kriossocc24}.

\subsubsection{Sustainability} 
Despite developments in sustainable computing, future distributed systems will continue to be considerable consumers of computing resources, leading to a significant energy consumption and carbon footprint \cite{imran2024towards, goldverg2025towards}. 
While energy consumption and the corresponding carbon emissions will be lower at the local/device level, it will drastically increase as we move along the tiers dimension, peaking at the Sky level. 
Specifically, the increasing use of specialized accelerators will reflect a shift toward matching workload demands with hardware that delivers higher performance per watt~\cite{hennessy2019new}. Near-data and in-memory processing techniques will help to reduce the substantial energy cost associated with moving data between memory, storage, and compute units~\cite{ghose2019processing}. Hardware systems will be designed to be more energy efficient, but increased energy efficiency at the hardware level will also lead to increased use of these computing resources and will eventually cause more energy consumption (Jevons' paradox~\cite{alcott2005jevons}). Hence, we envision that these advances alone will not be sufficient to keep up with the increased computing demands. More fundamentally, the distributed system infrastructures across the entire periodic space must evolve to become grid-aware and adaptive, coordinating compute, storage, and networking resources in response to the fluctuating capacity of the local power grids and availability of renewable energy~\cite{rodrigues2025carbon}.

We expect that sustainability will no longer be treated as a secondary objective in distributed systems. It will instead become an essential design principle, shaping decisions at every layer of the distributed computing stack. This requires the ability to measure, quantify, and ultimately reduce the energy consumption and carbon emissions at each of these layers. 
Hence, a key research opportunity lies in rethinking distributed systems to explicitly handle trade-offs among performance, energy consumption, and carbon emissions.


\subsubsection{Operational and Infrastructure Costs} 
Extrapolating observed trends across existing tiers, we anticipate that infrastructure cost per unit of compute (CapEx) 
and operational expenditure (OpEx) 
will continue to decrease as systems evolves into the Sky infrastructures. 
At first glance, deploying computing infrastructure in Sky domains, incorporating conventional on-Earth data centers, supercomputers, orbital, lunar, interplanetary, and quantum systems may appear prohibitively expensive. However, we anticipate that, over time, both CapEx and OpEx at these tiers may fall below conventional data centers. Particularly, as the technology of space-based data centers matures, the transportation and deployment cost of computing infrastructure will decrease \cite{kang2025cost} (e.g., via reusable rockets) 
and they can bypass the severe land, power, and water constraints on the earth. Sky computing will further benefit from abundant extra-planetary energy sources, such as solar, which can be utilized without many of the terrestrial constraints \cite{mankins2014case}. 
The Sky systems are inherently designed for autonomy and minimal human intervention, significantly reducing maintenance overhead in the long run. Nevertheless, thermal management is likely to remain a fundamental challenge, as cooling in space and on other planetary bodies is constrained by environmental conditions and limited heat dissipation mechanisms. 
However, while radiative cooling remains a fundamental limitation in space, the vacuum of space allows for passive thermal management \cite{starcloud_website}. 
Proponents such as Starcloud and Lonestar suggest that the long-term goal is for space-based, solar-powered data centers to have lower operating costs, potentially 97\% lower than on Earth once launch costs drop \cite{greengard2025}.

\subsection{Trends Towards Agent Abstraction}

\subsubsection{Democratization} 
\label{sec:democratization}
While distributed and cloud computing has permeated nearly every domain of life, its programming has remained exclusive to a small elite of highly specialized organizations \cite{wistc25}. We envision that developments in the abstraction dimension will reverse this trajectory. Computing tiers will evolve into cognitive partners equipped with AI-guided (agentic) programming abstractions \cite{hosseini2025role} that can model device-to-Sky continuum as one uniform compute-base and offer features such as self-composing workflows \cite{esashi2026action} and declarative deployment \cite{wisipdps26}. These will collapse the barrier between intent and implementation, and ultimately, make application development as accessible as writing a document.
\subsubsection{Agentic Systems}
We envision that Agentic systems will dominate the use of distributed infrastructure across all tiers--from device with a single agent, to other tiers (edge-to-Sky) with multiple agents interacting. The complexity of agent ``behaviors'' may range from: (i) Well-defined, state-oriented behaviors which are triggered through external events or agent priorities. In this instance, agent actions are constrained and limited by the type and range of events they observe to trigger such actions \cite{colledanchise2018behavior}. (ii) Dynamically updated behaviors which may be defined using domain-specific LLMs \cite{li2023api}. Such behaviors may also be composed dynamically, taking account of capacity and expertise of multiple agents. An example of (ii) is the use of either domain-specific or general purpose LLM models to characterize the agent's behavior. 

With agents behaving non-deterministically, understanding their interactions to solve complex/unknown tasks will become a challenge~\cite{schick2023toolformer}. The multi-agent coordination strategies may involve pre-defined interactions between agents, or may be dynamically developed by {\it discovering} the ``expertise'' of an agent \cite{ghafarollahi2025sciagents}, and the use of an LLM (agent) routing mechanism that is able to re-formulate and distribute a request across agents until it is accomplished. We envisage to see more domain-specific LLMs emerge that enable agent behaviors to be characterized (and differentiated) clearly.  

\subsubsection{Security and Trustworthiness} 
While higher abstraction improves system reliability through standardization and managed services, there is concern that agent abstraction introduces unprecedented security challenges that may degrade trustworthiness~\cite{bookchapterAkhil26}. Unlike traditional applications with deterministic execution paths, agents operate autonomously by planning, reasoning, and executing actions across the device-to-Sky continuum with minimal human oversight. This autonomy can expand the attack surface, as agents interact with external APIs, access sensitive data across organizational boundaries, and execute complex, multi-step workflows that are difficult to audit or constrain by conventional access control policies \cite{deng2025ai,he2025emerged}. Organizations cannot reliably foresee agents' behavior prior to deployment, and emergent capabilities may introduce vulnerabilities, such as prompt injection attacks \cite{owasp2025}, data exfiltration via tool misuse \cite{zhan2024injecagent}, or privilege escalation across distributed infrastructure \cite{ji2026taming}. Ensuring trustworthiness will require new security paradigms, including runtime behavioral monitoring \cite{aws2025_scoping}, sandboxed execution environments with fine-grained permission models \cite{google2025_gke, nvidia2025_sandbox}, and interpretable decision-making frameworks that expose agent reasoning chains for audit \cite{liang2025enhancing, ibm2025_transparency}. 

\subsubsection{Operational and Infrastructure Costs} 
We anticipate that both the OpEx and CapEx of agents and agentic AI services will be higher than those of application abstraction. Traditional applications rely on multiple abstraction layers in the software stack but are predominantly executed on CPUs and conventional computing resources. On one hand, agentic AI systems require substantial computational resources for both model training and inference, resulting in increased infrastructure and operational costs. On the other hand, such agents are expected to significantly reduce the cost of application development, deployment, and maintenance, which, in the long run, is likely to outweigh the additional
resource expenditures. 

\subsubsection{Governance} 

Agentic AI systems operate autonomously across distributed and heterogeneous resources, often spanning multiple owners and administrative domains while making real-time decisions that affect users, applications, and infrastructure. These systems commonly adopt a human-on-the-loop model \cite{fischer2021hotl}, where human intervention occurs only when necessary. As indicated by the periodic space, higher levels of abstraction delegate increasing control and operational responsibility to the underlying infrastructure. Thus, at the Agent abstraction level, governance is essential to ensure accountability, fairness, safety, and compliance despite decentralized control and partial trust among stakeholders. Approaches such as multi-level value alignment \cite{zeng2025multi} aim to keep autonomous decisions aligned with human, organizational, and societal objectives. Addressing these challenges requires scalable monitoring, data-driven insights \cite{donta2023}, and adaptive governance mechanisms that integrate policy enforcement, explainable decision-making, and sustainability-aware resource management to enable responsible deployment of agentic AI systems.

\section{Conclusions}
\label{sec:conc}
Inspired from the popular chemical periodic table, in this visional paper, we proposed a periodic space based on a continuum of computing tiers and abstraction levels that can capture the behavior of various solutions in the distributed computing landscape. The periodic space also has the ability to anticipate the behavior of emerging solutions in the landscape.
We used the periodic space to characterize the noticeable trends in the landscape. Given these trends, we anticipate structural changes will happen to the distributed computing landscape, and when considering the changes under periodic space, we can see strong forces to push innovation toward the high-abstraction, high-tier area.

\section*{Acknowledgments}
We would like to thank Dr. Mahadev Satyanarayanan, at Carnegie Melon University for his advice and mentoring throughout this study. Also, we appreciate feedback and comments from Dr. Christopher Stewart, at Ohio State University, and Dr. Alexandru Iosup, at Vrije Universiteit Amsterdam. This research is supported by the National Science Foundation (NSF) through CNS CAREER Award\# 2419588.

\bibliographystyle{IEEEtran}

\bibliography{references}

@article{ghose2019processing,
  title={Processing-in-memory: A workload-driven perspective},
  author={Ghose, Saugata and Boroumand, Amirali and Kim, Jeremie S and G{\'o}mez-Luna, Juan and Mutlu, Onur},
  journal={IBM Journal of Research and Development},
  volume={63},
  number={6},
  pages={3--1},
  year={2019},
  publisher={IBM}
}

@article{hennessy2019new,
  title={A new golden age for computer architecture},
  author={Hennessy, John L and Patterson, David A},
  journal={Communications of the ACM},
  volume={62},
  number={2},
  pages={48--60},
  year={2019},
  publisher={ACM New York, NY, USA}
}

@inproceedings{imran2024towards,
  title={Towards sustainable cloud software systems through energy-aware code smell refactoring},
  author={Imran, Asif and Kosar, Tevfik and Zola, Jaroslaw and Bulut, M Fatih},
  booktitle={Proceedings of the IEEE 17th International Conference on Cloud Computing},
  series = {CLOUD '24},
  pages={223--234},
  year={2024},
  organization={IEEE}
}

@misc{raspberrypi,
  author       = {{Raspberry Pi}},
  title        = {Raspberry {P}i},
  year         = {2026},
  url          = {https://www.raspberrypi.com/},
  note         = {Accessed: 2026-02-12}
}

@misc{nvidia_jetson_nano,
  author       = {{NVIDIA Corporation}},
  title        = {{Jetson Nano} Developer Kit},
  year         = {2026},
  url          = {https://developer.nvidia.com/embedded/jetson-nano},
  note         = {Accessed: 2026-02-12}
}

@online{knative,
  author  = {{Knative Authors}},
  title   = {Knative Official Website},
  year    = {2026},
  url     = {https://knative.dev/},
  urldate = {2026-02-12}
}

@online{aws_step_functions,
  author  = {{Amazon Web Services, Inc.}},
  title   = {{AWS Step Functions}},
  year    = {2026},
  url     = {https://aws.amazon.com/step-functions/},
  urldate = {2026-02-12}
}

@article{goldverg2025towards,
  title={Towards Carbon-Aware Data Transfers},
  author={Goldverg, Jacob and Jamil, Hasibul and Rodrigues, Elvis and Kosar, Tevfik},
  journal={IEEE Internet Computing},
  year={2025},
  publisher={IEEE}, 
  volume={29},
number={2},
pages={19-26},
}

@article{rodrigues2025carbon,
  title={Carbon-Aware Temporal Data Transfer Scheduling Across Cloud Datacenters},
  author={Rodrigues, Elvis and Goldverg, Jacob and Kosar, Tevfik},
  journal={Proceedings of the IEEE 18th International Conference on Cloud Computing},
  year={2025}, 
  series = {CLOUD '25},
}

@article{alcott2005jevons,
  title={Jevons' paradox},
  author={Alcott, Blake},
  journal={Ecological Economics},
  volume={54},
  number={1},
  pages={9--21},
  year={2005},
  publisher={Elsevier}
}

@inproceedings{mokhtari2024heet,
  title={{HEET}: A Performance Measure to Quantify Heterogeneity in Distributed Computing Systems},
  author={Mokhtari, Ali and Ghafouri, Saeid and Jamshidi, Pooyan and Salehi, Mohsen Amini},
  booktitle={Proceedings of the 17th  IEEE/ACM International Conference on Utility and Cloud Computing},
  series = {UCC '24},
  pages={17--26},
  year={2024},
}

@INPROCEEDINGS{felare23,
  author={Mokhtari, Ali and Hossen, Md Abir and Jamshidi, Pooyan and Salehi, Mohsen Amini},
  booktitle={Proceedings of the 15th IEEE International Conference on Cloud Computing}, 
  title={{FELARE}: Fair Scheduling of Machine Learning Tasks on Heterogeneous Edge Systems}, 
  year={2022},
  series = {CLOUD '22},
  pages={459-468},
  }

@INPROCEEDINGS{Akbari2025,
  author={Akbari, Negin and Grundy, John and Cheema, Aamir and Toosi, Adel N.},
  booktitle={Proceedings of the 2025 IEEE International Conference on Web Services}, 
  series = {ICWS '25},
  title={{IntentContinuum}: Using {LLMs} to Support Intent-Based Computing Across the Compute Continuum}, 
  year={2025},
  pages={573-583},
  doi={10.1109/ICWS67624.2025.00079}}

@article{esashi2026action,
  title={Action Engine: Automatic Workflow Generation in {FaaS}},
  author={Esashi, Akiharu and Lertpongrujikorn, Pawissanutt and Kato, Shinji and Salehi, Mohsen Amini},
  journal={Future Generation Computer Systems},
  volume={174},
  pages={107947},
  year={2026},
  publisher={Elsevier}
}

@inproceedings{CANLMAki26,
  author    = {Esashi, Akiharu and Lertpongrujikorn, Pawissanutt and Makino, Justin and Fujimoto, Yuibi and Amini Salehi, Mohsen},
  title     = {{Foundation CAN LM}: A Pretrained Language Model for Automotive {CAN} Data},
  booktitle = {Proceedings of the 37th IEEE Intelligent Vehicles Symposium},
  year      = {2026},
  month     = {Jun.},
  address   = {Detroit, MI, USA},
}

@ARTICLE{sky24,
  author={Trinh, Phuc V. and Sugiura, Shinya},
  journal={IEEE Communications Magazine}, 
  title={Quantum {I}nternet in the Sky: Vision, Challenges, Solutions, and Future Directions}, 
  year={2024},
  volume={62},
  number={10},
  pages={62-68},
}

@article{grabowska2024quantum,
  title={On quantum computing for artificial superintelligence},
  author={Grabowska, Anna and Gunia, Artur},
  journal={European Journal for Philosophy of Science},
  volume={14},
  number={2},
  pages={25},
  year={2024},
}

@inproceedings{kometsocc24,
author = {Pfandzelter, Tobias and Bermbach, David},
title = {Komet: A Serverless Platform for Low-Earth Orbit Edge Services},
booktitle = {Proceedings of the 2024 ACM Symposium on Cloud Computing},
year = {2024},
pages = {866–882},
numpages = {17},
location = {Redmond, WA, USA},
series = {SoCC '24}
}

@inproceedings{kriossocc24,
author = {Bhosale, Vaibhav and Gavrilovska, Ada and Bhardwaj, Ketan},
title = {Krios: Scheduling Abstractions and Mechanisms for Enabling a {LEO} Compute Cloud},
year = {2024},
booktitle = {Proceedings of the ACM Symposium on Cloud Computing},
pages = {322–340},
location = {Redmond, WA, USA},
series = {SoCC '24}
}

@inproceedings{sky21,
author = {Stoica, Ion and Shenker, Scott},
title = {From cloud computing to sky computing},
year = {2021},
booktitle = {Proceedings of the Workshop on Hot Topics in Operating Systems},
pages = {26–32},
numpages = {7},
location = {Ann Arbor, Michigan},
series = {HotOS '21}
}

@article{schick2023toolformer,
  title={{Toolformer: Language models can teach themselves to use tools}},
  author={Schick, Timo and Dwivedi-Yu, Jane and Dess{\`\i}, Roberto and Raileanu, Roberta and Lomeli, Maria and Hambro, Eric and Zettlemoyer, Luke and Cancedda, Nicola and Scialom, Thomas},
  journal={Advances in Neural Information Processing Systems},
  volume={36},
  pages={68539--68551},
  year={2023}
}

@article{ghafarollahi2025sciagents,
  title={{SciAgents}: Automating scientific discovery through bioinspired multi-agent intelligent graph reasoning},
  author={Ghafarollahi, Alireza and Buehler, Markus J},
  journal={Advanced Materials},
  volume={37},
  number={22},
  pages={2413523},
  year={2025},
  publisher={Wiley Online Library}
}

@article{li2023api,
  title={{Api-Bank: A comprehensive benchmark for tool-augmented LLMs}},
  author={Li, Minghao and Zhao, Yingxiu and Yu, Bowen and Song, Feifan and Li, Hangyu and Yu, Haiyang and Li, Zhoujun and Huang, Fei and Li, Yongbin},
  journal={arXiv preprint arXiv:2304.08244},
  year={2023}
}

@book{colledanchise2018behavior,
  title={Behavior Trees in Robotics and AI: An Introduction},
  author={Colledanchise, Michele and {\"O}gren, Petter},
  year={2018},
  publisher={CRC Press}
}

@article{hosseini2025role,
  title={The role of agentic {AI} in shaping a smart future: A systematic review},
  author={Hosseini, Soodeh and Seilani, Hossein},
  journal={Array},
  pages={100399},
  year={2025},
  publisher={Elsevier}
}

@article{bookchapterAkhil26,
  title={Downsides of Smartness Across Edge-Cloud Continuum in Modern Industry},
  author={Chigullapally, Akhil Gupta and Vittala, Sharvan and Hussian, Razin Farhan and Salehi, Mohsen Amini},
  journal={arXiv preprint arXiv:2603.29289},
  year={2026}
}

@INPROCEEDINGS{jewcloud24,
  author={Manatura, Sorawit and Chanikaphon, Thanawat and Chantrapornchai, Chantana and Amini Salehi, Mohsen},
  booktitle={Proceedings of the 17th IEEE International Conference on Cloud Computing}, 
  series = {CLOUD '24},
  title={{FastMig: Leveraging FastFreeze to Establish Robust Service Liquidity in Cloud 2.0}}, 
  year={2024},
  pages={81-90},
location={ShenZhen, China}
}

@article{black2010review,
  author  = {Black, David V.},
  title   = {Review of \emph{The Periodic Table: Its Story and Its Significance}},
  journal = {Technology and Culture},
  volume  = {51},
  number  = {1},
  pages   = {287--289},
  year    = {2010}
}

@inproceedings{hotmobile19,
author = {Satyanarayanan, Mahadev and Gao, Wei and Lucia, Brandon},
title = {The Computing Landscape of the 21st Century},
year = {2019},
pages = {45–50},
location = {Santa Cruz, CA, USA},
booktitle = {Proceedings of the 20th International Workshop on Mobile Computing Systems and Applications}, 
series = {HotMobile '19},
}

@inproceedings{bashir25,
author = {Ngata, Wacuka M and Bashir, Noman and Westerlaken, Michelle and Liote, Laurent and Chandio, Yasra and Olivetti, Elsa},
title = {The Cloud Next Door: Investigating the Environmental and Socioeconomic Strain of Datacenters on Local Communities},
year = {2025},
booktitle = {Proceedings of the ACM SIGCAS/SIGCHI Conference on Computing and Sustainable Societies},
pages = {769–774},
series = {COMPASS '25}
}

@inproceedings{wissocc24,
author = {Lertpongrujikorn, Pawissanutt and Nguyen, Hai Duc and Salehi, Mohsen Amini},
title = {Streamlining Cloud-Native Application Development and Deployment with Robust Encapsulation},
year = {2024},
booktitle = {Proceedings of the ACM Symposium on Cloud Computing},
pages = {847–865},
numpages = {19},
location = {Redmond, WA, USA},
series = {SoCC '24}
}

@inproceedings{wiscloud23,
  title={{Object as a service ({OaaS})}: Enabling object abstraction in serverless clouds},
  author={Lertpongrujikorn, Pawissanutt and Salehi, Mohsen Amini},
  booktitle={Proceedings of the 16th IEEE International Conference on Cloud Computing},
  pages={238--248},
  year={2023},
  location={Chicago},
  series={CLOUD '23}
}

@article{liu2018space,
  title={Space-air-ground integrated network: A survey},
  author={Liu, Jiajia and Shi, Yongpeng and Fadlullah, Zubair Md and Kato, Nei},
  journal={IEEE Communications Surveys \& Tutorials},
  volume={20},
  number={4},
  pages={2714--2741},
  year={2018},
  publisher={IEEE}
}

@book{moris2021infrastructure,
  author = {Morris, K.},
  title={Infrastructure as Code: Designing and Delivering Dynamic Systems for the Cloud Age},
  edition = {3},
  year={2025},
  publisher={O’Reilly Media}
}

@article{jonas2019cloud,
  title={Cloud programming simplified: A {B}erkeley view on serverless computing},
  author={Jonas, Eric and Schleier-Smith, Johann and Sreekanti, Vikram and Tsai, Chia-Che and Khandelwal, Anurag and Pu, Qifan and Shankar, Vaishaal and Carreira, Joao and Krauth, Karl and Yadwadkar, Neeraja and others},
  journal={arXiv preprint arXiv:1902.03383},
  year={2019}
}

@inproceedings{denby2020orbital,
  title={Orbital edge computing: Nanosatellite constellations as a new class of computer system},
  author={Denby, Bradley and Lucia, Brandon},
  booktitle={Proceedings of the 25th International Conference on Architectural Support for Programming Languages and Operating Systems},
  series = {ASPLOS '20}, 
  pages={939--954},
  year={2020}
}

@inproceedings{charyyev2020latency,
  title={Latency comparison of cloud datacenters and edge servers},
  author={Charyyev, Batyr and Arslan, Engin and Gunes, Mehmet Hadi},
  booktitle={Proceedings of the 2020 IEEE Global Communications Conference},
  series = {GLOBECOM '20},
  pages={1--6},
  year={2020},
  organization={IEEE}
}

@ARTICLE{wistc25,
  author={Lertpongrujikorn, Pawissanutt and Salehi, Mohsen Amini},
  journal={IEEE Transactions on Computers}, 
  title={Object as a Service: Simplifying Cloud-Native Development through Serverless Object Abstraction}, 
  year={2026},
  volume={75},
  number={2},
  pages={423-434},
}

@inproceedings{satya25,
  author={Satyanarayanan, Mahadev and Dong, Qifei and Xu, Jingao and Pillai, Padmanabhan},
  title={Towards Mobile {AI} That Is Accurate and Fast},
  booktitle={Proceedings of the 7th IEEE International Conference on Cognitive Machine Intelligence},
  year={2025},
  series= {COGMI '25},
  month={Nov.}
}

@inproceedings{ma2023network,
  title={Network characteristics of leo satellite constellations: A starlink-based measurement from end users},
  author={Ma, Sami and Chou, Yi Ching and Zhao, Haoyuan and Chen, Long and Ma, Xiaoqiang and Liu, Jiangchuan},
  booktitle={IEEE INFOCOM 2023-IEEE Conference on Computer Communications},
  pages={1--10},
  year={2023},
  organization={IEEE}
}

@article{Armbrust2010Cloud,
  title={A view of cloud computing},
  author={Armbrust, Michael and Fox, Armando and Griffith, Rean and Joseph, Anthony D and Katz, Randy and Konwinski, Andy and Lee, Gunho and Patterson, David and Rabkin, Ariel and Stoica, Ion and others},
  journal={Communications of the ACM},
  volume={53},
  number={4},
  pages={50--58},
  year={2010},
  publisher={ACM New York, NY, USA}
}

@article{Atzori2010IoT,
  title={The internet of things: A survey},
  author={Atzori, Luigi and Iera, Antonio and Morabito, Giacomo},
  journal={Computer networks},
  volume={54},
  number={15},
  pages={2787--2805},
  year={2010},
  publisher={Elsevier}
}

@techreport{WhiteHouse2020CET,
  author      = {{The White House}},
  title       = {National Strategy for Critical and Emerging Technologies},
  institution = {Executive Office of the President},
  year        = {2020},
  month       = oct,
  address     = {Washington, DC},
  url         = {https://trumpwhitehouse.archives.gov/wp-content/uploads/2020/10/National-Strategy-for-CET.pdf}
}

@techreport{EC2020Digital,
  author      = {{European Commission}},
  title       = {Shaping {Europe}'s Digital Future},
  institution = {European Commission},
  year        = {2020},
  month       = feb,
  number      = {COM(2020) 67 final},
  address     = {Brussels},
  url         = {https://eur-lex.europa.eu/legal-content/en/TXT/?uri=CELEX:52020DC0067}
}

@article{Duan2023DAI,
  title={Distributed artificial intelligence empowered by end-edge-cloud computing: A survey},
  author={Duan, Sijing and Wang, Dan and Ren, Ju and Lyu, Feng and Zhang, Ye and Wu, Huaqing and Shen, Xuemin},
  journal={IEEE Communications Surveys \& Tutorials},
  volume={25},
  number={1},
  pages={591--624},
  year={2022},
  publisher={IEEE}
}

@article{Bommasani2021Foundation,
  author    = {Bommasani, Rishi and Hudson, Drew A. and Adeli, Ehsan 
               and Altman, Russ and Arora, Simran and von Arx, Sydney 
               and Bernstein, Michael S. and Bohg, Jeannette 
               and Bosselut, Antoine and Brunskill, Emma 
               and others},
  title     = {On the Opportunities and Risks of Foundation Models},
  journal   = {arXiv preprint arXiv:2108.07258},
  year      = {2022},
  doi       = {10.48550/arXiv.2108.07258}
}

@article{Theis2017MooresLaw,
  title={The end of {M}oore's law: A new beginning for information technology},
  author={Theis, Thomas N and Wong, H-S Philip},
  journal={Computing in Science \& Engineering},
  volume={19},
  number={2},
  pages={41--50},
  year={2017},
  publisher={IEEE}
}

@techreport{UNCTAD2019Digital,
  author      = {{United Nations Conference on Trade and Development}},
  title       = {Digital Economy Report 2019: Value Creation and Capture: 
                 Implications for Developing Countries},
  institution = {United Nations},
  year        = {2019},
  address     = {Geneva},
  isbn        = {978-92-1-112955-7},
  url         = {https://unctad.org/publication/digital-economy-report-2019}
}

@book{WorldBank2016WDR,
  author    = {{World Bank}},
  title     = {World Development Report 2016: Digital Dividends},
  year      = {2016},
  publisher = {World Bank},
  address   = {Washington, DC},
  doi       = {10.1596/978-1-4648-0671-1},
  isbn      = {978-1-4648-0671-1}
}

@article{Thompson2021Decline,
  title={The decline of computers as a general purpose technology},
  author={Thompson, Neil C and Spanuth, Svenja},
  journal={Communications of the ACM},
  volume={64},
  number={3},
  pages={64--72},
  year={2021},
  publisher={ACM New York, NY, USA}
}

@article{ji2026taming,
  title={Taming Various Privilege Escalation in {LLM}-Based Agent Systems: A Mandatory Access Control Framework},
  author={Ji, Zimo and Wu, Daoyuan and Jiang, Wenyuan and Ma, Pingchuan and Li, Zongjie and Gao, Yudong and Wang, Shuai and Li, Yingjiu},
  journal={arXiv preprint arXiv:2601.11893},
  year={2026}
}

@article{zhan2024injecagent,
  title={{InjecAgent}: Benchmarking indirect prompt injections in tool-integrated large language model agents},
  author={Zhan, Qiusi and Liang, Zhixiang and Ying, Zifan and Kang, Daniel},
  journal={arXiv preprint arXiv:2403.02691},
  year={2024}
}

@article{he2025emerged,
  title={The emerged security and privacy of {LLM} agent: A survey with case studies},
  author={He, Feng and Zhu, Tianqing and Ye, Dayong and Liu, Bo and Zhou, Wanlei and Yu, Philip S},
  journal={ACM Computing Surveys},
  volume={58},
  number={6},
  pages={1--36},
  year={2025},
  publisher={ACM New York, NY}
}

@article{deng2025ai,
  title={{AI} agents under threat: A survey of key security challenges and future pathways},
  author={Deng, Zehang and Guo, Yongjian and Han, Changzhou and Ma, Wanlun and Xiong, Junwu and Wen, Sheng and Xiang, Yang},
  journal={ACM Computing Surveys},
  volume={57},
  number={7},
  pages={1--36},
  year={2025},
}

@inproceedings{wisipdps26,
  author={Lertpongrujikorn, Pawissanutt and Kwon, Juahn and Nguyen, Hai Duc and Amini Salehi, Mohsen},
  title={{EdgeWeaver}: Accelerating {IoT} Application Development Across Edge-Cloud Continuum},
  booktitle={Proceedings of the 40th IEEE International Parallel and Distributed Processing Symposium},
 series = {IPDPS '26},
 year={2026},
 month={May.},
location={New Orleans, USA}
}

@INPROCEEDINGS{globecome23,
  author={Chanikaphon, Thanawat and Salehi, Mohsen Amini},
  booktitle={Proceedings of the IEEE Global Communications Conference}, 
  title={{UMS}: Live Migration of Containerized Services across Autonomous Computing Systems}, 
  year={2023},
  pages={467-472},
}

@misc{owasp2025,
  author = {{OWASP Foundation}},
  title = {{LLM01}: 2025 Prompt Injection},
  year = {2025},
  howpublished = {OWASP Gen AI Security Project},
  url = {https://genai.owasp.org/llmrisk/llm01-prompt-injection}
}

@misc{aws2025_scoping,
  author = {Brown, Aaron and Saner, Matt},
  title = {The Agentic {AI} Security Scoping Matrix: A framework for securing autonomous {AI} systems},
  year = {2025},
  month = {November},
  howpublished = {AWS Security Blog},
  url = {https://aws.amazon.com/it/blogs/security/the-agentic-ai-security-scoping-matrix-a-framework-for-securing-autonomous-ai-systems/},
}

@misc{google2025_gke,
  author = {Brandon Royal},
  title = {Introducing Agent sandbox: Strong guardrails for agentic {AI on Kubernetes and GKE}},
  year = {2025},
  month = {November},
  howpublished = {Google Cloud Blog},
  url = {https://cloud.google.com/blog/products/containers-kubernetes/agentic-ai-on-kubernetes-and-gke},
}

@misc{nvidia2025_sandbox,
  author = {Rich Harang},
  title = {Practical Security Guidance for Sandboxing Agentic Workflows and Managing Execution Risk},
  year = {2026},
  month = {January},
  howpublished = {NVIDIA Technical Blog},
}

@article{liang2025enhancing,
  title={Enhancing performance of explainable {AI} models with constrained concept refinement},
  author={Liang, Geyu and Michielssen, Senne and Fattahi, Salar},
  journal={arXiv preprint arXiv:2502.06775},
  year={2025}
}

@misc{ibm2025_transparency,
  author = {Jonker, Alexandra and Gomstyn, Alice and McGrath, Amanda},
  title = {What is {AI} transparency?},
  year = {2025},
  month = {November},
  howpublished = {IBM Think Topics}
}

@article{lagunas2024leo,
	author = {Lagunas, Eva and Chatzinotas, Symeon and Ottersten, Bj{\"o}rn},
	doi = {10.1038/s44287-024-00088-9},
	isbn = {2948-1201},
	journal = {Nature Reviews Electrical Engineering},
	number = {10},
	pages = {656--665},
	title = {Low-Earth orbit satellite constellations for global communication network connectivity},
	volume = {1},
	year = {2024},
}

@inproceedings{nguyen2025efficient,
    author = {Nguyen, Hai Duc and Chien, Andrew A},
    title = {Efficient Performance Guarantees for Function-as-a-Service with Cloud Allocators},
    year = {2025},
    doi = {10.1145/3721462.3730948},
    booktitle = {Proceedings of the 26th International Middleware Conference},
    pages = {99–113},
    series = {Middleware '25},
}

@inproceedings{nguyen2019real,
  title={Real-time serverless: Enabling application performance guarantees},
  author={Nguyen, Hai Duc and Zhang, Chaojie and Xiao, Zhujun and Chien, Andrew A},
  booktitle={Proceedings of the 5th International Workshop on Serverless Computing},
  pages={1--6},
  year={2019}
}

@online{microsoft_office,
  author  = {{Microsoft Corporation}},
  title   = {Microsoft Office},
  year    = {2026},
  url     = {https://www.office.com/},
  urldate = {2026-02-12}
}

@inproceedings{park2023generative,
  title={Generative agents: Interactive simulacra of human behavior},
  author={Park, Joon Sung and O'Brien, Joseph and Cai, Carrie Jun and Morris, Meredith Ringel and Liang, Percy and Bernstein, Michael S},
  booktitle={Proceedings of the 36th annual acm symposium on user interface software and technology},
  pages={1--22},
  year={2023}
}

@article{fu2025autonomous,
  title={Autonomous data agents: A new opportunity for smart data},
  author={Fu, Yanjie and Wang, Dongjie and Ying, Wangyang and Wang, Xinyuan and Zhang, Xiangliang and Liu, Huan and Pei, Jian},
  journal={arXiv preprint arXiv:2509.18710},
  year={2025}
}

@article{bird2022taking,
  title={Taking Flight with {C}opilot: Early insights and opportunities of AI-powered pair-programming tools},
  author={Bird, Christian and Ford, Denae and Zimmermann, Thomas and Forsgren, Nicole and Kalliamvakou, Eirini and Lowdermilk, Travis and Gazit, Idan},
  journal={Queue},
  volume={20},
  number={6},
  pages={35--57},
  year={2022},
  publisher={ACM New York, NY, USA}
}

@article{li2024survey,
  title={A survey on {LLM}-based multi-agent systems: Workflow, infrastructure, and challenges},
  author={Li, Xinyi and Wang, Sai and Zeng, Siqi and Wu, Yu and Yang, Yi},
  journal={Vicinagearth},
  volume={1},
  number={1},
  pages={9},
  year={2024},
  publisher={Springer}
}

@article{satyanarayanan2017emergence,
  title={The emergence of edge computing},
  author={Satyanarayanan, Mahadev},
  journal={Computer},
  volume={50},
  number={1},
  pages={30--39},
  year={2017},
  publisher={IEEE}
}

@article{rallis2025interfacing,
  title={Interfacing quantum computing systems with high-performance computing systems: an overview},
  author={Rallis, Konstantinos and Liliopoulos, Ioannis and Varsamis, Georgios D and Tsipas, Evangelos and Karafyllidis, Ioannis G and Sirakoulis, Georgios Ch and Dimitrakis, Panagiotis},
  journal={arXiv preprint arXiv:2509.06205},
  year={2025}
}

@article{kumari2025quantum,
  title={Quantum Cloud Computing: Key Technologies, Challenges, and Opportunities},
  author={Kumari, Anisha and Behera, Ranjan Kumar and Sahoo, Bibhudatta},
  journal={Advances in Quantum Inspired Artificial Intelligence: Techniques and Applications},
  pages={99--124},
  year={2025},
  publisher={Springer}
}

@inproceedings{younis2024comprehensive,
  title={A comprehensive analysis of cloud service models: {IaaS, PaaS, and SaaS} in the context of emerging technologies and trend},
  author={Younis, Rehmana and Iqbal, Mansoor and Munir, Khalid and Javed, Muhammad Aaqib and Haris, Muhammad and Alahmari, Saad},
  booktitle={Proceedings of the 2024 International Conference on Electrical, Communication and Computer Engineering},
  pages={1--6},
  year={2024},
  organization={IEEE}
}

@article{burns2016borg,
  title={{Borg, Omega, and Kubernetes}},
  author={Burns, Brendan and Grant, Brian and Oppenheimer, David and Brewer, Eric and Wilkes, John},
  journal={Communications of the ACM},
  volume={59},
  number={5},
  pages={50--57},
  year={2016},
  publisher={ACM New York, NY, USA}
}

@article{kang2025cost,
  title={Cost Effectiveness of Reusable Launch Vehicles Depending on the Payload Capacity},
  author={Kang, Si-Yoon and Jo, Min-Seon and Choi, Jeong-Yeol and Yang, Soo Seok},
  journal={Aerospace},
  volume={12},
  number={5},
  pages={364},
  year={2025},
  publisher={MDPI}
}

@book{mankins2014case,
  title={The Case for Space Solar Power},
  author={Mankins, John},
  year={2014},
  publisher={Virginia Edition Publishing Houston, TX, USA}
}

@online{starcloud_website,
  author  = {{Starcloud}},
  title   = {Starcloud Official Website},
  year    = {2026},
  url     = {https://www.starcloud.com/},
  urldate = {2026-02-12}
}

@inproceedings{nicolae2024diaspora,
  title={Diaspora: Resilience-enabling services for real-time distributed workflows},
  author={Nicolae, Bogdan and Wozniak, Justin M and Bicer, Tekin and Nguyen, Hai and Patel, Parth and Pan, Haochen and Gueroudji, Amal and Gonthier, Maxime and Hayot-Sasson, Valerie and Huerta, Eliu and others},
  booktitle={2024 IEEE 20th International Conference on e-Science},
  series = {e-Science '24},
  pages={1--9},
  year={2024},
}

@article{nguyen2025resilient,
  title={Resilient execution of distributed X-ray image analysis workflows},
  author={Nguyen, Hai Duc and Bicer, Tekin and Nicolae, Bogdan and Kettimuthu, Rajkumar and Huerta, Eliu A and Foster, Ian T},
  journal={Frontiers in High Performance Computing},
  volume={3},
  pages={1550855},
  year={2025},
  publisher={Frontiers Media SA}
}

@article{nain2022towards,
  title={Towards edge computing in intelligent manufacturing: Past, present and future},
  author={Nain, Garima and Pattanaik, KK and Sharma, GK},
  journal={Journal of Manufacturing Systems},
  volume={62},
  pages={588--611},
  year={2022},
  publisher={Elsevier}
}

@inproceedings{huynh2017deepmon,
  title={Deepmon: Mobile gpu-based deep learning framework for continuous vision applications},
  author={Huynh, Loc N and Lee, Youngki and Balan, Rajesh Krishna},
  booktitle={Proceedings of the 15th Annual International Conference on Mobile Systems, Applications, and Services},
  pages={82--95},
  year={2017}
}

@inproceedings{yeon2025watchhar,
  title={WatchHAR: Real-time On-device Human Activity Recognition System for Smartwatches},
  author={Yeon, Taeyoung and Xu, Vasco and Hoffmann, Henry and Ahuja, Karan},
  booktitle={Proceedings of the 27th International Conference on Multimodal Interaction},
  pages={387--394},
  year={2025}
}

@article{zhang2025research,
  title={Research on webassembly runtimes: A survey},
  author={Zhang, Yixuan and Liu, Mugeng and Wang, Haoyu and Ma, Yun and Huang, Gang and Liu, Xuanzhe},
  journal={ACM Transactions on Software Engineering and Methodology},
  volume={34},
  number={8},
  pages={1--47},
  year={2025},
  publisher={ACM New York, NY}
}

@article{shalf2020,
    author = {Shalf, John},
    title = {The future of computing beyond {Moore}'s Law},
    journal = {Philosophical Transactions of the Royal Society A: Mathematical, Physical and Engineering Sciences},
    volume = {378},
    number = {2166},
    pages = {20190061},
    year = {2020},
    doi = {10.1098/rsta.2019.0061},
}

@article{donta2023,
	author = {Donta, Praveen Kumar and Sedlak, Boris and Casamayor Pujol, Victor and Dustdar, Schahram},
	doi = {10.1186/s40537-023-00737-0},
	journal = {Journal of Big Data},
	number = {1},
	pages = {53},
	title = {Governance and sustainability of distributed continuum systems: A big data approach},
	volume = {10},
	year = {2023},
}

@article{fischer2021hotl,
author = {Fischer, Joel E. and Greenhalgh, Chris and Jiang, Wenchao and Ramchurn, Sarvapali D. and Wu, Feng and Rodden, Tom},
title = {In-the-loop or on-the-loop? {I}nteractional arrangements to support team coordination with a planning agent},
journal = {Concurrency and Computation: Practice and Experience},
volume = {33},
number = {8},
pages = {e4082},
doi = {https://doi.org/10.1002/cpe.4082},
year = {2021}
}

@article{kounev2023serverless,
author = {Kounev, Samuel and Herbst, Nikolas and Abad, Cristina L. and Iosup, Alexandru and Foster, Ian and Shenoy, Prashant and Rana, Omer and Chien, Andrew A.},
title = {Serverless Computing: What It Is, and What It Is Not?},
year = {2023},
volume = {66},
number = {9},
doi = {10.1145/3587249},
journal = {Communications of the ACM},
pages = {80–92},
}

@inproceedings{haas2017wasm,
author = {Haas, Andreas and Rossberg, Andreas and Schuff, Derek L. and Titzer, Ben L. and Holman, Michael and Gohman, Dan and Wagner, Luke and Zakai, Alon and Bastien, JF},
title = {Bringing the web up to speed with {WebAssembly}},
year = {2017},
doi = {10.1145/3062341.3062363},
booktitle = {Proceedings of the 38th ACM SIGPLAN Conference on Programming Language Design and Implementation},
series = {PLDI '17},
pages = {185–200},
}

@inproceedings{menetrey2022wasm,
    author = {M\'{e}n\'{e}trey, J\"{a}mes and Pasin, Marcelo and Felber, Pascal and Schiavoni, Valerio},
    title = {WebAssembly as a Common Layer for the Cloud-edge Continuum},
    year = {2022},
    doi = {10.1145/3526059.3533618},
    booktitle = {Proceedings of the 2nd Workshop on Flexible Resource and Application Management on the Edge (FRAME '22)},
    pages = {3–8},
}

@INPROCEEDINGS{kakati2024wasm,
  author={Kakati, Sangeeta and Brorsson, Mats},
  booktitle={Proceedings of the IEEE 24th International Symposium on Cluster, Cloud and Internet Computing}, 
  series = {CCGrid '24},
  title={A Cross-Architecture Evaluation of {WebAssembly} in the Cloud-Edge Continuum}, 
  year={2024},
  pages={337-346},
  doi={10.1109/CCGrid59990.2024.00046}
}

@article{zeng2025multi,
    title={Multi-level Value Alignment in Agentic {AI} Systems: Survey and Perspectives}, 
    author={Wei Zeng and Hengshu Zhu and Chuan Qin and Han Wu and Yihang Cheng and Sirui Zhang and Xiaowei Jin and Yinuo Shen and Zhenxing Wang and Feimin Zhong and Hui Xiong},
    year={2025},
    journal={arXiv preprint arXiv:2506.09656},
}

@MISC{greengard2025,
  author={Samuel Greengard},
  title={Datacenters Go to Space},
  year = {2025}, 
  journal = {Communications of the ACM}, 
  howpublished={https://cacm.acm.org/news/datacenters-go-to-space/}
}

\end{document}